\documentclass[aps,a4paper,twocolumn,nofootinbib,superscriptaddress]{revtex4}
\usepackage{amsmath,amssymb,graphicx,color,dsfont}
\usepackage{pdfpages}
\pdfoutput=1

\newcommand{\ket}[1]{\ensuremath{\left| #1 \right\rangle}}

\newcommand{\omegam}{\ensuremath{\omega_\textrm{M}}}

\newcommand{\XL}{\ensuremath{X_\textrm{L}}}
\newcommand{\PL}{\ensuremath{P_\textrm{L}}}
\newcommand{\XM}{\ensuremath{X_\textrm{M}}}

\newcommand{\X}{\ensuremath{\widetilde{X}}}
\newcommand{\Y}{\ensuremath{\widetilde{Y}}}
\newcommand{\Xp}{\ensuremath{\widetilde{X}^{\phi}}}
\newcommand{\Yp}{\ensuremath{\widetilde{Y}^{\phi}}}
\newcommand{\XX}{\ensuremath{\widetilde{P}}}
\newcommand{\YY}{\ensuremath{\widetilde{Q}}}

\newcommand{\PLout}{\ensuremath{P_\textrm{L}^{\textrm{out}}}}
\newcommand{\PLin}{\ensuremath{P_\textrm{L}^{\textrm{in}}}}
\newcommand{\XLout}{\ensuremath{X_\textrm{L}^{\textrm{out}}}}
\newcommand{\XLin}{\ensuremath{X_\textrm{L}^{\textrm{in}}}}

\newcommand{\etal}{\emph{et al}.}

\newcommand{\APL}[4]{#4. \textit{Appl. Phys. Lett.} \textbf{#1,} #2 (#3)}

\newcommand{\CRP}[4]{#4. \textit{C. R. Physique} \textbf{#1,} #2 (#3)}
\newcommand{\CQG}[4]{#4. \textit{Class. Quantum Grav.} \textbf{#1,} #2 (#3)}

\newcommand{\Nature}[4]{#4. \textit{Nature} \textbf{#1,} #2 (#3)}
\newcommand{\NatPhys}[4]{#4. \textit{Nature Physics} \textbf{#1,} #2 (#3)}
\newcommand{\NatPhot}[4]{#4. \textit{Nature Photonics} \textbf{#1,} #2 (#3)}
\newcommand{\NatComm}[4]{#4. \textit{Nat. Commun.} \textbf{#1,} #2 (#3)}
\newcommand{\NJP}[4]{#4. \textit{New J. Phys.} \textbf{#1,} #2 (#3)}

\newcommand{\PNAS}[4]{#4. \textit{Proc. Natl. Acad. Sci. USA} \textbf{#1,} #2 (#3)}
\newcommand{\PRL}[4]{#4. \textit{Phys. Rev. Lett.} \textbf{#1,} #2 (#3)}
\newcommand{\PRA}[4]{#4. \textit{Phys. Rev. A} \textbf{#1,} #2 (#3)}
\newcommand{\PRB}[4]{#4. \textit{Phys. Rev. B} \textbf{#1,} #2 (#3)}

\newcommand{\PRD}[4]{#4. \textit{Phys. Rev. D} \textbf{#1,} #2 (#3)}
\newcommand{\PRX}[4]{#4. \textit{Phys. Rev. X} \textbf{#1,} #2 (#3)}
\newcommand{\RMP}[4]{#4. \textit{Rev. Mod. Phys.} \textbf{#1,} #2 (#3)}
\newcommand{\Science}[4]{#4. \textit{Science} \textbf{#1,} #2 (#3)}

\begin{document}

\title{
Non-linear optomechanical measurement of mechanical motion
}

\author{G. A. Brawley}
\thanks{These authors contributed equally to this work.}
\affiliation{ARC Centre for Engineered Quantum Systems, School of Mathematics and Physics, The University of Queensland, Brisbane, Queensland 4072, Australia}

\author{M. R. Vanner\footnote{Email correspondence: michael.vanner@physics.ox.ac.uk}}
\thanks{These authors contributed equally to this work.}
\affiliation{ARC Centre for Engineered Quantum Systems, School of Mathematics and Physics, The University of Queensland, Brisbane, Queensland 4072, Australia}
\affiliation{Clarendon Laboratory, Department of Physics, University of Oxford, OX1 3PU, United Kingdom}

\author{P. E. Larsen}
\affiliation{Department of Micro- and Nanotechnology,
Technical University of Denmark, DTU Nanotech,
DK-2800 Kongens Lyngby, Denmark}

\author{S. Schmid}
\affiliation{Department of Micro- and Nanotechnology,
Technical University of Denmark, DTU Nanotech,
DK-2800 Kongens Lyngby, Denmark}

\author{A. Boisen}
\affiliation{Department of Micro- and Nanotechnology,
Technical University of Denmark, DTU Nanotech,
DK-2800 Kongens Lyngby, Denmark}

\author{W. P. Bowen}
\affiliation{ARC Centre for Engineered Quantum Systems, School of Mathematics and Physics, The University of Queensland, Brisbane, Queensland 4072, Australia}

\date{\today}


\begin{abstract}
Precision measurement of non-linear observables is an important goal in all facets of quantum optics. This allows measurement-based non-classical state preparation, which has been applied to great success in various physical systems, and provides a route for quantum information processing with otherwise linear interactions. In cavity optomechanics much progress has been made using linear interactions and measurement, but observation of non-linear mechanical degrees-of-freedom remains outstanding. Here we report the observation of displacement-squared thermal motion of a micro-mechanical resonator by exploiting the intrinsic non-linearity of the radiation pressure interaction. Using this measurement we generate bimodal mechanical states of motion with separations and feature sizes well below 100~pm. Future improvements to this approach will allow the preparation of quantum superposition states, which can be used to experimentally explore collapse models of the wavefunction and the potential for mechanical-resonator-based quantum information and metrology applications.
\end{abstract}

\maketitle




A key tool in quantum optics is the use of measurement to conditionally prepare quantum states. This technique, often simply referred to as `conditioning', has been applied to generate non-Gaussian quantum states for confined microwave fields~\cite{Deleglise2008}, travelling optical fields~\cite{Ourjoumtsev2007, Bimbard2010}, and superconducting systems~\cite{Riste2013}. In addition, quantum measurements are of vital importance to many quantum computation protocols~\cite{KLM2001}. In cavity optomechanics, light circulating inside an optical resonator is used to manipulate and measure the motion of a mechanical element via the radiation-pressure interaction~\cite{Aspelmeyer2013}. After an optomechanical interaction performing a measurement on the light can then be used to conditionally prepare mechanical states of motion. Such mechanical conditioning has been performed with measurements of the mechanical position~\cite{Vanner2011, Vanner2013, Szorkovszky2013}, however, thus far, conditioning has not been performed with a measurement of a non-linear mechanical degree of freedom. One exciting prospect of such a non-linear measurement is the detection of phonon number jumps~\cite{Thompson2008,Miao2009}, thus demonstrating mechanical energy quantisation. Here, we experimentally demonstrate non-Gaussian conditional state preparation of a mechanical resonator by performing position-squared measurements. Our approach does not rely on a dispersive coupling to the mechanical position squared~\cite{Thompson2008} but instead utilises the optical non-linearity of the radiation pressure coupling as was proposed in Ref.~\cite{VannerPRX2011}. Such a position-squared measurement can ultimately be used to prepare a mechanical superposition state~\cite{Jacobs2009} as the measurement does not obtain sign information of the displacement. Studying the dynamics of these states can be used to test models of decoherence beyond standard quantum mechanics~\cite{Ghirardi1986, Diosi1989, Penrose1996, Kleckner2008, RomeroIsart2011, Blencowe2013, VannerPRX2011} and for the development of mechanical quantum sensors.

\begin{figure}
\begin{center}
\includegraphics[width=1\hsize]{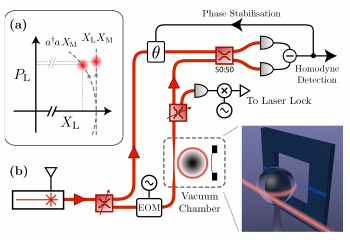}
\end{center}
\caption{
\textbf{Concept and experimental apparatus.} \textbf{(a)} The intrinsic optical non-linearity of an optomechanical interaction gives rise to rotations of the optical field in phase-space that can be observed in both the phase (\PL) and amplitude (\XL) quadratures. Conventionally this interaction is linearised in the weak coupling regime, leading to optical phase quadrature displacements only. Our optical setup \textbf{(b)} can measure an arbitrary optical quadrature of light from the optomechanical system using homodyne interferometry and is capable of observing the higher-order terms in displacement, described by Eq.~(\ref{Eq:XPout}). The optomechanical system consists of a nanostring mechanical resonator evanescently coupled to an optical microsphere resonator \textbf{(c)} (not shown to scale), which is mounted in a high vacuum chamber ($<10^{-6}$mbar). The drive laser is stabilised to the cavity resonance using the Pound Drever Hall technique. Polarization control is not shown for clarity. 
}


\label{Fig:Setup}
\end{figure}

The intra-cavity Hamiltonian for such an optomechanical system in a frame rotating at the optical carrier frequency is 
$H/\hbar = \omegam b^\dagger b - g_0 a^\dagger a (b+b^\dagger)$
where $a$ ($b$) is the optical (mechanical) annihilation operator, $\omegam$ is the mechanical angular frequency, and $g_0$ is the zero-point optomechanical coupling rate. Quite generally, optomechanics experiments to-date have focused on dynamics describable by a linearised model of the radiation pressure interaction~\cite{Aspelmeyer2013}, where the photon number operator is approximated by $a^\dagger a\simeq N+\sqrt{N}(a^\dagger+a)$ and $N$ is the mean intracavity photon number. In this approximation, mechanical displacements give rise to displacements of the optical phase quadrature leaving the optical amplitude quadrature is unchanged. Fundamentally however, the radiation-pressure interaction is non-linear~\cite{VannerPRX2011,Borkje2013}, and generates mechanical position dependent rotations of the intracavity optical field, as is illustrated in Fig. \ref{Fig:Setup}(a). For small changes in the mechanical position, the optical phase quadrature changes linearly in proportion to the  mechanical displacement, and the optical amplitude quadrature reduces in proportion to the mechanical displacement squared. By choosing which optical quadrature to observe with homodyne detection, one may selectively measure the mechanical displacement or displacement-squared~\cite{VannerPRX2011}. Since a displacement-squared measurement does not distinguish between positive and negative displacement, mechanical superposition states may be prepared by measurement~\cite{Jacobs2009}. A necessary requirement for the optical interaction to effect a direct measurement of the displacement-squared is that the mechanical motion is negligible during the intra-cavity photon lifetime, i.e. the bad cavity regime ($\kappa \gg \omegam$, where $\kappa$ is the optical amplitude decay rate of the cavity). This should be contrasted to other approaches operating in the resolved sideband regime ($\kappa \ll \omegam$), where cavity-averaged displacement-squared interactions have been predicted to allow the observation of the mechanical phonon-number~\cite{Thompson2008,Miao2009}. In this work we observe optomechanical dynamics arising from the non-linearity of the radiation-pressure interaction and, utilizing this non-linearity, perform non-Gaussian state generation by measurement of displacement-squared mechanical motion.

In the bad-cavity regime the intracavity field can be approximated as $a\simeq \sqrt{N}/(1-i\lambda\XM) + \xi$, where $\lambda = \sqrt{2}g_0/\kappa$ quantifies the optomechanical interaction strength, $\xi$ is the intracavity noise term and ${\XM=(b+b^\dagger)/\sqrt{2}}$ is the mechanical displacement in units of the mechanical quantum noise. Taylor expanding this expression, the time-dependent optical output quadratures are then
\begin{equation}
\label{Eq:XPout}
\begin{split}
\XLout & = 2\sqrt{\kappa N}\left[1 - \lambda^2\XM^2 + ...\right] - \XLin,\\
\PLout & = 2\sqrt{\kappa N}\left[\lambda\XM - \lambda^3\XM^3 + ...\right] - \PLin,
\end{split}
\end{equation}
where $\XL = (a + a^\dagger)/\sqrt{2}$, and $\PL = i(a^\dagger - a)/\sqrt{2}$. 
Conventionally, experimental optomechanics has focused on the leading, linear term in the expansion of the phase quadrature. In this linearised picture, only a single spectral peak at the mechanical resonance frequency is expected. Higher-order terms in mechanical displacement however, give rise to spectral peaks at the respective multiples of the mechanical resonance frequency which are only described by the full nonlinear optomechanical Hamiltonian. Precision measurement of these higher-order terms enables the conditional preparation of non-Gaussian states which, in a quantum regime, produces highly non-classical
states~\cite{Hudson}. Here we use the quadratic term to conditionally prepare classical bimodal states from an initial room-temperature thermal state of mechanical motion. 

Conditional state preparation can be understood as a combination of Bayesian inference, i.e. updating our knowledge of the system, and back-action. (See the supplementary information for a more detailed discussion.) Subsequent measurements following the conditioning step can be used to characterize the state prepared. Conditioning can be applied to a continuous measurement regime, where the sequence of measurement outcomes, known as the measurement record, give the conditional state evolution, or quantum trajectory. We utilise such a continuous measurement here and we would like to highlight that for this type of measurement, initialisation of the mechanical oscillator near its ground state is not required to generate non-classical mechanical states. This insensitivity to initial thermal occupation is because a continuous position-squared measurement also serves to purify the state.


\begin{figure}
\includegraphics[width=1.0\hsize]{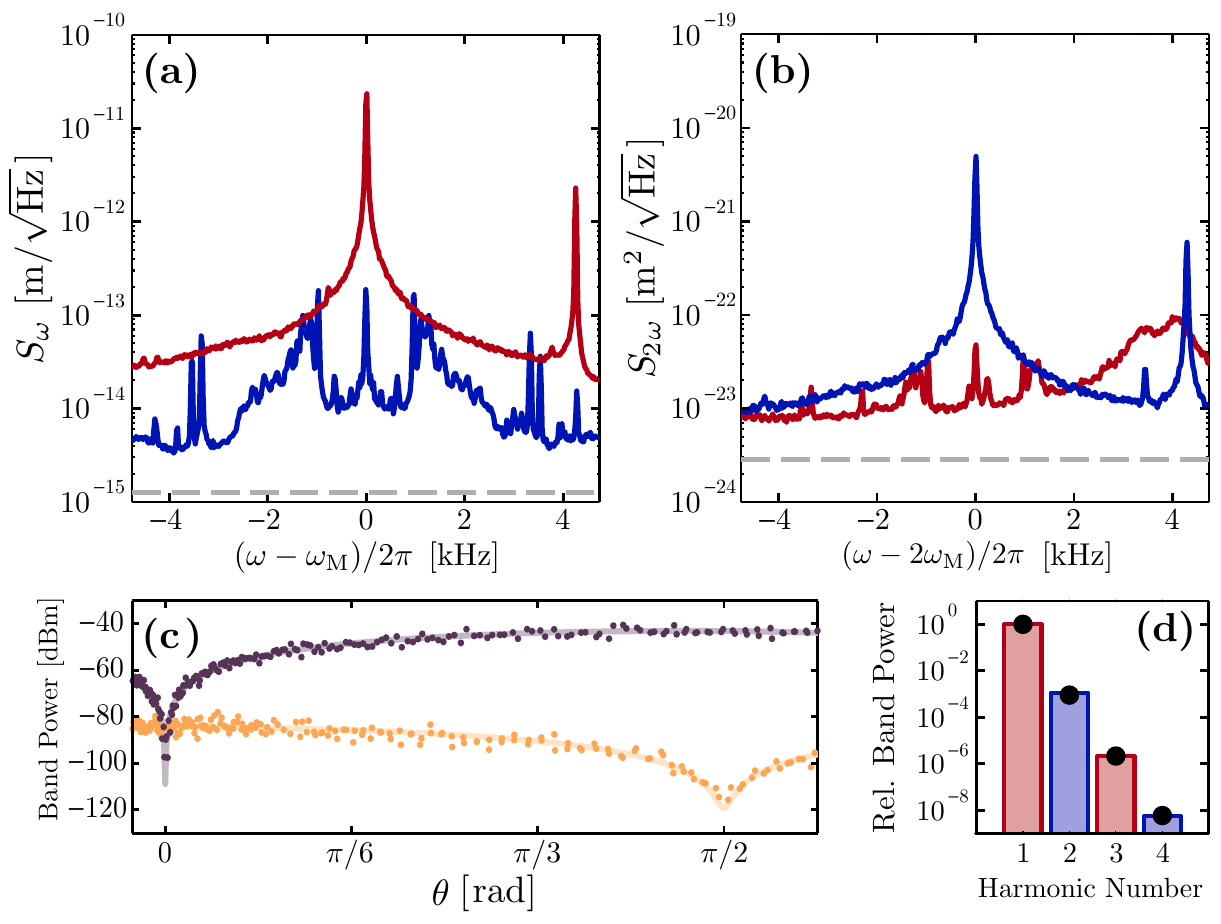}
\caption{
\textbf{Observation of linear and quadratic motion of a mechanical oscillator.} \textbf{(a)} Measurement of the optical phase quadrature (red trace) at the mechanical frequency, \omegam, shows a Lorentzian mechanical displacement spectrum, which is strongly suppressed when measuring the optical amplitude quadrature (blue trace). Sidebands appear in the amplitude quadrature measurement due to mix-up of low frequency noise.
\textbf{(b)} Measurement of the optical amplitude quadrature at $2\omegam$ (blue trace), shows the Lorentzian mechanical displacement squared noise, which is suppressed when measuring the optical phase quadrature (red trace). Note the second flexural mode of the string, located within a few kHz of $2\omegam$, is not transduced due to the positioning of the microsphere. In (a-b) the optical shot noise is shown as a grey dashed line, using a measurement bandwidth of $20$~Hz.
\textbf{(c)} The optical noise power measured over a  $51$~Hz bandwidth at $\omegam$ (purple) and $2\omegam$ (yellow) as a function of the optical homodyne angle. \textbf{(d)} The relative observed powers in each of the mechanical harmonics when measuring the optical phase (red) and amplitude quadratures (blue); bars - theory; dots - experimental data. Note the error bars in the power measurements are smaller than the dot size.
}
\label{Fig:SpectraPlots}
\end{figure}

\section*{Results}

A schematic of our non-linear optomechanics experiment is shown in Fig.~\ref{Fig:Setup}(b). We use a near-field cavity optomechanical setup~\cite{Anetsberger2009}, where a mechanical SiN nanostring oscillator~\cite{Schmid2011} is placed in close proximity to a 60~$\mu$m diameter optical microsphere resonator and interacts with the optical cavity field via the optical evanescent field (Fig.~\ref{Fig:Setup}(c)). The nanostring has dimensions $1000 \times 10 \times 0.054~\mu$m (length$\times$width$\times$thickness) and a fundamental mechanical resonance frequency of $\omegam/2\pi = 100.2$~kHz. From the known dimensions and density we estimate an effective mass of $m = 0.86$~ng. A continuous-wave fibre laser, operating at 1,559~nm, is locked on resonance with a whispering-gallery mode of the microsphere. We measure an optical amplitude decay rate of $\kappa/2\pi = 25.6$~MHz and mechanical linewidth of $\gamma/2\pi = 0.7$~Hz. An evanescent optomechanical coupling of $7.6$~MHz/nm was determined (see methods), corresponding to a coupling rate of $g_0/2\pi=75$~Hz. We use approximately 2~$\mu$W of optical drive power resulting in an intracavity photon number $N = 2.4\times10^4$.
A fibre-based Mach-Zehnder interferometer is used to perform homodyne detection and thereby selectively measure a quadrature of the optical output field. 

Fig.~\ref{Fig:SpectraPlots}(a) and (b) show the observed homodyne noise power spectra for both optical phase and amplitude quadratures at the mechanical frequency and the second harmonic, respectively. At $\omegam$ (Fig.~\ref{Fig:SpectraPlots}(a)) we observe a Lorentzian peak in the phase quadrature from the thermal motion of the oscillator, which corresponds to an RMS displacement of $124$~pm corresponding to a thermal occupation of $\bar{n}\simeq 10^8$. The thermal noise is resolved with $85$~dB of signal relative to the homodyne noise power when the signal is blocked, which corresponds to an ideal displacement sensitivity of $1.3\times 10^{-15}$~m$/\sqrt{\textrm{Hz}}$. In practice, the signal beam is not shot noise limited due to cavity, acoustic, and laser noise which raise the measurement imprecision by roughly an order of magnitude. By setting the interferometer phase to measure the optical amplitude quadrature, the linear measurement of mechanical motion is suppressed by approximately $45$~dB. At this quadrature, information about the displacement-squared mechanical motion is observed in a frequency band centred at $2\omegam$ (Fig.~\ref{Fig:SpectraPlots}(b)). We observe a Lorentzian peak with a linewidth of $1.5$~Hz, which to within the measurement uncertainty, is equal to twice the linewidth at \omegam. The signal-to-noise at this frequency is 65~dB relative to the homodyne noise, which corresponds to a calibrated ideal displacement-squared sensitivity of $3.3\times 10^{-24}~\textrm m^2/\sqrt{\textrm{Hz}}$. 

Fig.~\ref{Fig:SpectraPlots}(c) shows the band power in the first and second harmonics as a function of the interferometer phase. The powers in each band are expected to follow sine and cosine squared functions (fitted). The observed suppression of the linear measurement allows an upper bound to be placed on the phase instability of the cavity and interferometer locks of at most $5\times10^{-3}$~rad.
Fig.~\ref{Fig:SpectraPlots}(d) shows the observed relative noise powers up to the $4^{\textrm{th}}$ harmonic of the mechanical frequency. 
The expected noise powers can be computed with the Isserlis-Wick theorem, showing excellent agreement with experiment.

Of primary interest in this work is the lowest order nonlinear measurement term in the optical amplitude quadrature, proportional to $\XM^2$. To describe this quantitatively we introduce the slowly varying quadratures of motion, $X$ and $Y$, defined via $\XM(t) = X(t)\cos\omegam t + Y(t)\sin\omegam t$. The mechanical displacement-squared signal can then be written as
\begin{equation}
\label{Eq:xm2DCSinCos}
\begin{split}
\XM^2=~&{\textstyle\frac{1}{2}}(X^2+Y^2)~+~{\textstyle\frac{1}{2}}(X^2-Y^2)\cos(2\omegam t)\\
& +~{\textstyle\frac{1}{2}}(XY+YX)\sin(2\omegam t)
\end{split}
\end{equation}
where for later convenience, the displacement-squared quadratures of motion are defined $P={\textstyle\frac{1}{2}}(X^2-Y^2)$ and $Q={\textstyle\frac{1}{2}}(XY+YX)$. By inspection, it can be seen that the quadratic measurement has spectral components both at DC and $2\omegam$. 
Higher order terms in the expansion Eq.(\ref{Eq:XPout}) can in principle contribute to the signal at $2\omegam$, however since $\lambda^2\bar{n}\ll 1$, the quadratic term is the only term to contribute substantial power at $2\omegam$. Consequently, linear and quadratic components of the measurement can be spectrally separated and therefore, at an appropriate homodyne angle, measured simultaneously.

\begin{figure}
\includegraphics[width=0.8\hsize]{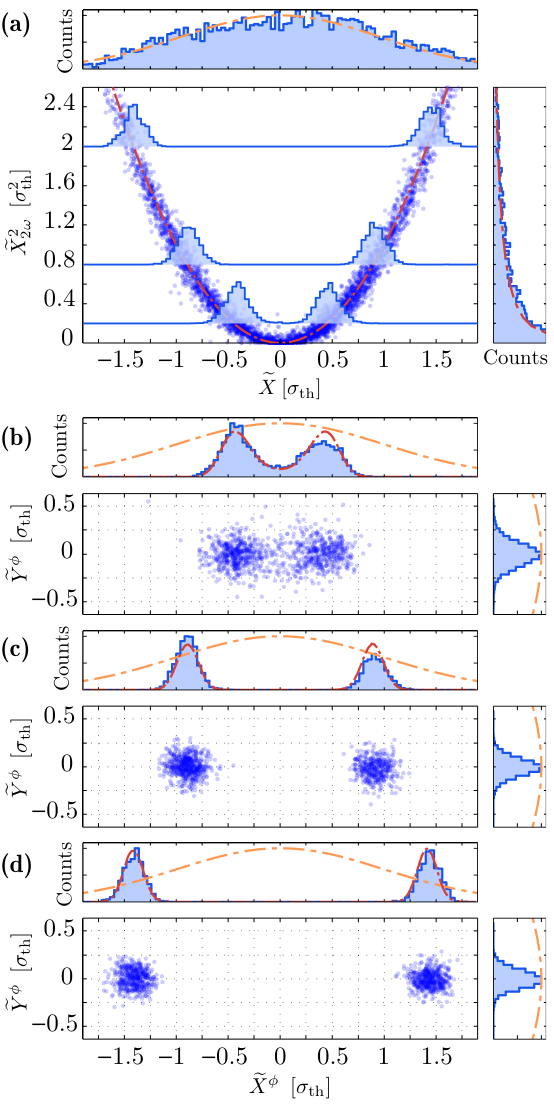}
\hfill
\caption{
\textbf{Bimodal state preparation via non-linear measurement.}
Using a homodyne angle of $\pi/4$, a high-fidelity measurement of both linear and quadratic motion of the mechanical oscillator at frequencies of $\omegam$ and $2\omegam$, respectively, is obtained.
In \textbf{(a)} the quadratic measurement outcomes ($\X_{2\omega}^2$) obtained from the $2\omega_\textrm{M}$ signal are plotted against the linear outcomes ($\X$) obtained from the signal at $\omega_\textrm{M}$. The histogram of $\X$ measurements (above) is well described by a Gaussian thermal distribution with standard deviation $\sigma_\textrm{th}=124$~pm, while the histogram of the $\X_{2\omega}^2$ measurements (right) forms a chi-squared distribution.
Figures \textbf{(b-d)} show the phase-space distributions (and associated histograms) of states conditionally prepared using data at $2\omegam$ and obtained using a read-out at $\omegam$.
The conditionally rotated read-out data is decomposed into conjugate quadratures, labelled $\Xp$ and $\Yp$. The chosen quadratic conditioning values are \textbf{(b)} ($2\XX^{2\phi}$ = 0.2), \textbf{(c)} ($2\XX^{2\phi}$ = 0.8), and \textbf{(d)} ($2\XX^{2\phi}$ = 2.0). The same quadratic conditioning values are indicated as overlay histograms in (a). The histograms in (b-d) are normalised to their peak value to more easily allow the width of the features to be compared to the initial thermal state (orange dash-dot curve).
The red curves overlayed in the histograms in $\Xp$ are determined via numerical simulation, with the signal-to-noise ratios for the linear and quadratic measurements as fitting parameters.
}
\label{Fig:Conditional}
\end{figure}

To perform both state preparation and state reconstruction we set a homodyne angle of $\pi/4$, which allows simultaneous high fidelity linear measurement (for state reconstruction) and quadratic measurements (for state preparation).
The photocurrent generated at the homodyne output is then digitized into 4 second blocks at $5$~MS/sec, which are filtered in post-processing at \omegam~and $2\omegam$~and decomposed into sine and cosine components to obtain the respective quadratures of motion in each frequency band.
For a large signal-to-noise ratio, the squares of each quadrature of motion can be estimated from the measurements of $P$ and $Q$ via the nonlinear transformations $X^2\simeq\X_{2\omega}^2=\sqrt{{\XX}^2+{\YY}^2}+{\XX}$ and $Y^2\simeq\Y_{2\omega}^2=\sqrt{{\XX}^2+{\YY}^2}-{\XX}$, where the tildes denote the (noise inclusive) measurement outcomes of the respective quantity. These transformations allow the recovery of a classical estimate of $\XM^2$ without knowledge of the signal at DC. Fig. \ref{Fig:Conditional}(a) plots the $\X_{2\omega}^2$ estimates thus obtained from the $2\omegam$ signal against the cosine mechanical position quadrature, $\X$, obtained from the measurement at \omegam. A clear quadratic relationship between the two measurements is observed, validating the displacement-squared nature of the $2\omegam$ peak.

Conditioning based upon the outcome of the quadratic measurement can be used to prepare non-Gaussian states. In the most basic approach, conditioning $\YY=0$ and $\XX=C$, for some constant $C$, will produce a bimodal state with a separation of $2\sqrt{2C}$. However, to make more efficient use of the available data, we additionally perform a mechanical phase rotation for each sample in the measurement record. Firstly, at each discrete sample we find a rotation by a phase angle $2\phi$ such that the new rotated variable $\YY^{2\phi}=\YY\cos(2\phi)-\XX\sin(2\phi)$ is equal to zero. As a result, in the frame rotated by the half angle, $\phi$, correlation between the two mechanical quadratures $\X$ and $\Y$ is conditionally eliminated.
Secondly, we condition on a particular magnitude of $\XX^{2\phi}=\XX\cos(2\phi)+\YY\sin(2\phi)$ (see methods). This operation localizes the phase space distribution of the reconstructed state to two small regions as shown in Fig.~\ref{Fig:Conditional}(b), with a separation dependent upon the conditioning value. These states, although classical, are evidently bimodal and non-Gaussian.
Extending this protocol to a regime where the quadratic measurement rate dominates all decoherence processes, a macroscopic quantum superposition state can be generated. Indeed, as detailed later, a simulated state prepared in this way is shown in Fig. \ref{Fig:OpenQuantumSimulations}(b).

\section*{Discussion}

At present, in opto- and electro-mechanics, techniques towards measurement of non-linear observables of mechanical motion include coupling to two-level systems \cite{OConnell2010} and radiation-pressure interactions coupling to the displacement-squared, such as the `membrane-in-the-middle' (MiM) approach \cite{Thompson2008}. In the latter approach, a mechanically vibrating element is appropriately placed within an optical standing wave in a cavity to give a displacement-squared dispersive coupling of the form $H_{\textrm{int}}/\hbar = \mu_0 a^\dagger a (b+b^\dagger)^2$, where $\mu_0$ is the zero-point quadratic coupling rate. This interaction, when operating in the resolved sideband regime, in principle allows for the observation of quantum jumps in the mechanical phonon number. Experiments exhibiting this type of coupling include dielectric membrane systems~\cite{Thompson2008, Sankey2010, Flowers2012}, trapped cold atoms~\cite{Purdy2010}, trapped microspheres~\cite{Li2011}, or double-disk structures~\cite{Lin2009}. However it should be noted that in these systems, quadratic coupling rates $\mu_0$ are typically orders of magnitude smaller than attainable linear coupling rates $g_0$.

In contrast to a displacement-squared Hamiltonian coupling, the scheme employed here gives an effective quadratic coupling rate of $g_0^2/\kappa$~\cite{VannerPRX2011}, which should be compared to $\mu_0$ defined above. For the modest linear coupling achieved in the present work, we have a $g_0^2/2\pi\kappa= 2.2\times 10^{-4}$~Hz. Crucially, since the coupling rate in our scheme scales as $g_0^2$, substantial gains are possible by improving $g_0$. For instance, the coupling rate for a state-of-the-art evanescently coupled nanostring-microcavity system as described in \cite{Anetsberger2011} is $g_0^2/2\pi\kappa \simeq 5\times 10^{-2}$ Hz and for a state-of-the-art electro-mechanical system \cite{Teufel2011} is $g_0^2/2\pi\kappa \simeq 2\times 10^{-1}$ Hz . Furthermore, other optical systems with exceptional linear coupling rates \cite{Amir2013}, should have quadratic coupling rates using our scheme as high as $160$~Hz.
In comparison, a state-of-the-art MiM system as described in Ref.~\cite{Flowers2012} has a quadratic coupling rate of $\mu_0/2\pi = 6.0 \times 10^{-3}$~Hz.
Thus a significantly larger effective quadratic coupling is possible using our protocol. Noteably, the quadratic measurement rate resulting from this fundamental coupling may be boosted by a coherent optical drive, which makes entering the quantum regime more feasible.


In all measurement-based quantum state preparation schemes, the measurement rate must dominate the sum of all decoherence processes due to coupling of the system to the environment. In MiM displacement-squared coupling protocols, even in a zero-temperature environment, this introduces the challenging requirement of single-photon strong coupling \mbox{($g_0/\kappa > 1$)~\cite{Miao2009}}. By contrast, our scheme offers a route to relax this stringent criterion. In our scheme, when state conditioning is performed with only the quadratic motion component of the detected signal, decoherence from the linearised radiation pressure noise on the mechanics precludes non-classical state generation outside of a single-photon strong coupling regime, similar to MiM. However, by including feed-back, this form of decoherence can, in the limit of perfect detection efficiency, be completely eliminated (see supplementary). This is because the amplitude quadrature measurement records not only the $X^2$ mechanical motion, but also the optical intracavity amplitude fluctuations near the mechanical resonance frequency. Since these fluctuations drive the linearised radiation pressure back-action on the mechanics, suitable feed-back to the motion of the mechanical oscillator can cancel this radiation pressure noise. Additionally, with this feed-back, the mechanical dynamics reduce to a similar form as with the displacement-squared dispersive Hamiltonian coupling. In the realistic case of imperfect detection efficiency $\eta$, the decoherence can be suppressed up to a factor of $1-\eta$. This results in the coupling strength requirement $g_0^2/\kappa^2 > (1- \eta)/2\eta$ to reach a quantum regime (see supplementary for derivation). For example, with a detection efficiency of $\eta=0.98$ the single photon coupling rate need only be one tenth of the strong coupling requirement. Additionally, the quadratic measurement rate must dominate rethermalisation, i.e. $4 \eta N g_0^4/\kappa^3 > \gamma (\bar n + 1/2)$. Provided the coupling strength criterion is satisfied, rethermalisation can be made insignificant with only modest intra-cavity photon numbers in cryogenic systems.

Based on these criteria, the quantum regime of our scheme can be achieved with current atom-optomechanical systems. For instance, quantum superposition states of motion could immediately be implemented with the approach in Ref.~\cite{Brennecke2008}, provided the detection efficiency exceeds $10\%$. Furthermore, solid-state optomechanical devices have seen rapid gains in performance over the past decade, with both optical and microwave systems now operating within three orders of magnitude of the single photon strong coupling regime. For example, an effective coupling rate of $g_0/\kappa = 0.04$ has recently been achieved in a superconducting microwave optomechanical device \cite{Pirkkalainen2015}, and with modest modifications, it is expected the system will approach the strong coupling regime. When $g_0/\kappa = 0.04$, the generation of quantum superposition states using our protocol requires detection efficiency on the order of $99.7\%$. However, with a one order of magnitude improvement in the coupling rate, the required detection efficiency drops to $76\%$, such that in combination with state of the art amplifiers \cite{Macklin2015}, nonclassical state generation using our protocol could be realised.

\begin{figure}
\hbox{\hspace{0.05\hsize}\includegraphics[width=1.0\hsize]{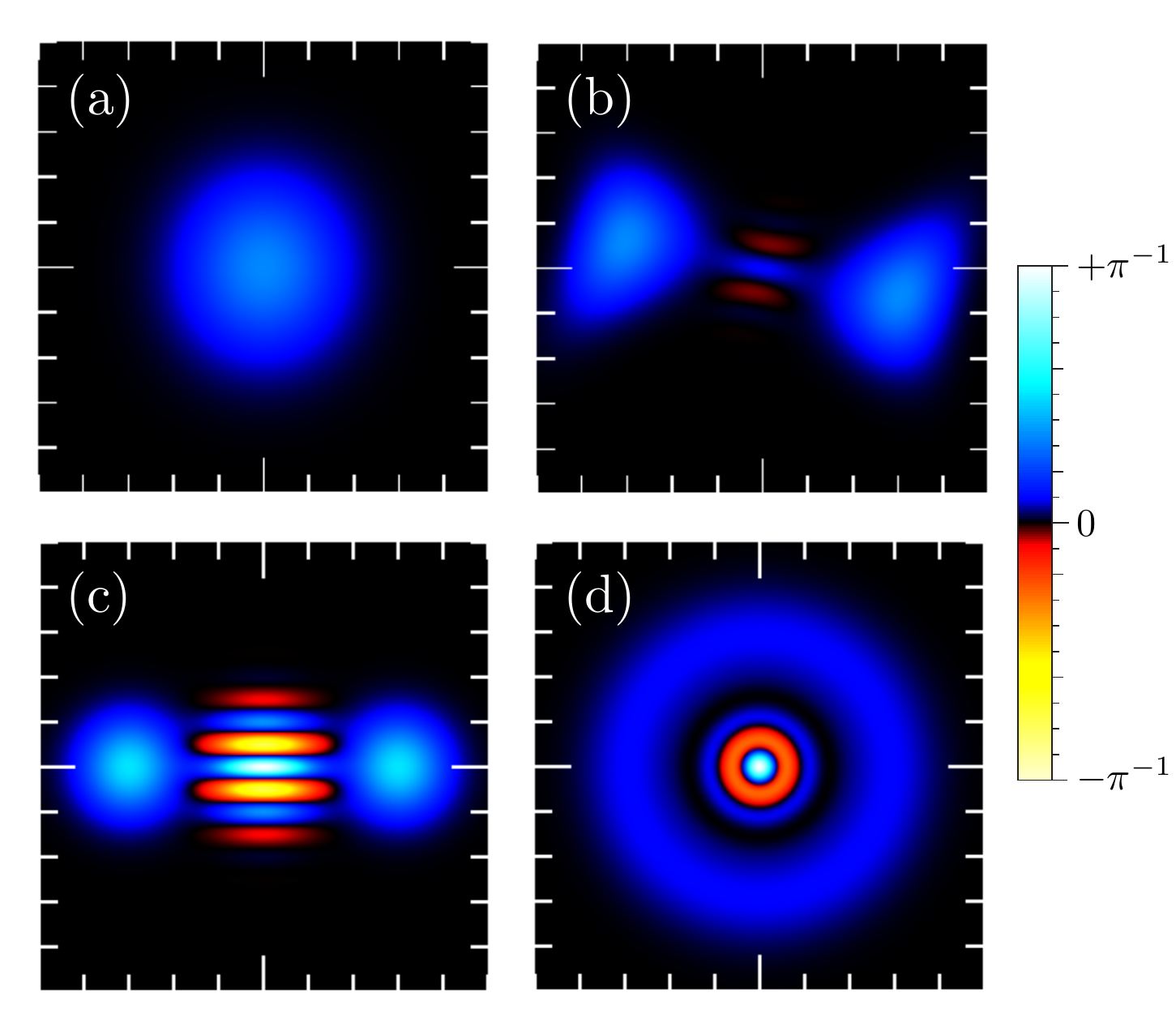}}
\caption{
\textbf{Open quantum system simulations.}
Figures \textbf{(a-b)} show the conditional state evolution of the mechanical oscillator under continuous measurement of the AC component of the $X^2$ signal, as computed from a master equation simulation. The initial state with $\bar{n}=1$ is shown in \textbf{(a)}. As the state evolves, negativity appears in the Wigner function. This is illustrated in \textbf{(b)} at time $t=6.4\times2\pi/\omegam$. This state may be compared to the canonical cat-sate of $\ket{\psi}=(\ket{3/\sqrt{2}}+\ket{-3/\sqrt{2}})/\sqrt{2}$, shown in \textbf{(c)}. An animation of the Wigner function time evolution at intermediate times is contained within the supplementary.
Figures \textbf{(c-d)} show the effect of phonon number decoherence on an initial even cat state. Notably after tracing over a strong phonon number measurement \textbf{(d)}, negativity is still preserved in the Wigner function. Further details of both the master equation simulation and analysis of the decoherence processes are contained in the supplementary information.
}
\label{Fig:OpenQuantumSimulations}
\end{figure}

Technical limitations may also play a r\^ole in the implementation of our protocol in a quantum regime. For example, fluctuations or offsets in the interferometer phase or cavity lock will result in linear coupling to the environment, and therefore an additional source of decoherence.
Linear coupling can also be introduced undesirably due to the presence of other mechanical modes in the system, which mix via the optical non-linearity with the mode of interest. Indeed the sum beat between the two mechanical modes in Fig.~\ref{Fig:SpectraPlots}(a) is observed as the $+4.3$~kHz peak in Fig.~\ref{Fig:SpectraPlots}(b).
These additional linear decoherence channels are expected to be negligible compared with the decoherence due to linear radiation pressure back-action as detailed in the supplementary.
The currently un-utilised DC component of the homodyne signal constitutes an additional technical decoherence channel. However, unlike the decoherence mechanisms discussed above, this channel is nonlinear, carrying information about phonon number rather than mechanical position. This can be seen from the expansion of the quadratic motion in terms of the creation and annihilation operators, $\XM^2=b^\dagger b + \frac{1}{2} + \frac{1}{2}(bb+b^\dagger b^\dagger)$, and identifying the number operator $n=b^\dagger b$ plus a constant as the DC part. As a result, loss of the DC information generates phase diffusion on the mechanical state. Somewhat strikingly, the non-classicality of states generated by $X^2$ measurement can in fact be quite robust against this form of decoherence. In the supplementary information we analyse phonon-number decoherence of an initial superposition state, showing that Wigner function negativity is preserved even in the presence of a complete loss of phonon number information to the environment. This result is illustrated in Fig.~\ref{Fig:OpenQuantumSimulations}~(c)~and~(d).

Finally, in order to elucidate the precise effect of the combination of all identified decoherence processes on the state conditioned via continuous quadratic measurement, a master equation simulation of our system was performed. The results for a particular trajectory are briefly summarised in the Wigner functions presented in Fig. \ref{Fig:OpenQuantumSimulations}(a-b). Shown in Fig. \ref{Fig:OpenQuantumSimulations}(a) is an initial thermal state of the mechanical oscillator. After a period of continuous measurement of the AC component of $\XM^2$, and in the presence of DC and thermal decoherence, this initially symmetric Gaussian state evolves into a non-Gaussian bimodal quantum state, exhibiting Wigner negativity near the origin, as shown in Fig. \ref{Fig:OpenQuantumSimulations}(b). Notably, qualitatively similar states have previously been shown to form in a different system under continuous position-squared measurement and conditioning \cite{Jacobs2009}. The states prepared by our protocol exhibit many of the properties of the canonical Schr\"odinger cat state of Fig. \ref{Fig:OpenQuantumSimulations}(c) and are highly non-classical. As a result, even in the presence of the identified decoherence mechanisms, we can conclude our protocol can give rise to interesting non-classical states.

To summarize, by exploiting the non-linearity inherent in the radiation pressure interaction, we report nonlinear measurement of thermo-mechanical motion in an optomechanical system.  Utilising the measurement of displacement-squared motion, we demonstrate the first measurement-based state preparation of mechanical non-Gaussian states. Furthermore, we propose a method using feed-back to extend this protocol to a quantum regime without requiring single-photon strong coupling.
Favourable scaling of the coupling rate in our approach makes realistic the possibility of observing the displacement-squared fluctuations of the mechanical ground state in the near future. With sufficiently high detection efficiency, this would allow for mechanical quantum superposition state preparation.
As a result, this experiment paves the way for quantum non-Gaussian state preparation of mechanical motion via measurement with applicability to a number of other physical systems, such as cold atoms~\cite{Brennecke2008}, atomic spin ensembles~\cite{Hammerer2010}, optomechanical systems~\cite{Amir2013} and superconducting microwave circuits~\cite{Hatridge2013, Murch2013, Teufel2011, Pirkkalainen2015}.

\section*{Methods}
\footnotesize{

\textbf{Linear Calibration Procedure.} We determine the evanescent optomechanical coupling by displacing the nanostring by a known distance using a piezoelectric element and measuring the resulting frequency shift on the optical resonance. The frequency shift is calibrated via modulation of known frequency applied to the laser. We then establish the response of the homodyne by sweeping the laser detuning over the optical resonance and measuring the slope of the phase response. This parameter combined with the previously determined optomechanical coupling rate gives the total response of the combined cavity interferometer system in [V/nm], allowing direct calibration of the time domain data in [nm]. We calibrate the response of our spectrum analyser by applying a test tone of known amplitude, which using the time domain calibration gives a spectral peak of known displacement spectral density.

\textbf{Quadratic Calibration Procedure.} Frequency domain calibration of the quadratic measurement is performed by ensuring the calibrated RMS displacement, obtained from the linear measurement, is consistent with the noise power of the $2\omegam$ peak, in accordance with the Isserlis-Wick theorem.
In the time domain, a simple regression is used between the square of the linear measurement (\X) and the quadratic measurement ($\X_{2\omega}^2$). We verify that these procedures are consistent, to within known uncertainties, with one another and with the value of $\lambda^2\bar{n}$ computed from the independently measured system parameters.

\textbf{State Conditioning.}
From the continuously acquired data, estimates of the quadratures at $2\omegam$ ($\omegam$) are obtained with the use of causal (acausal) decaying exponential filters, in order to time separate the conditioning and read-out phases. From the filtered data at each discrete time step, we rotate the vector \{\XX,\YY\} by an angle $2\phi$, such that a new vector $\{\XX^{2\phi},\YY^{2\phi}\}=\{(\XX^2 + \YY^2)^{\frac{1}{2}},0\} $ is obtained. The simultaneously acquired linear data \{\X,\Y\} is then rotated through the half angle, $\phi$, to obtain $\{\Xp,\Yp\}=\{\X\cos(\phi)+\Y\sin(\phi),\Y\cos(\phi)-\X\sin(\phi)\}$. 
For state preparation, the rotated linear data is conditioned upon the value of $\XX^{2\phi}$, which is proportional to $\frac{1}{2}(\Xp)^2$. We choose a conditioning window 4 times smaller than the quadratic measurement uncertainty. When the conditioning criterion is satisfied, the state is read-out using the rotated linear data $\{\X^{\phi},\Y^{\phi}\}$. All the data presented here have been generated from three 4 second blocks of sampled homodyne output.

}


 \section*{Acknowledgments}

We would like to thank K.~E.~Khosla, G.~J.~Milburn, and T.~M.~Stace for useful discussion. This research was supported primarily by the ARC CoE for Engineered Quantum Systems (CE110001013). M.R.V. acknowledges support provided by an ARC Discovery Project (DP140101638). P.E.L., S.S., and A. B. acknowledge funding from the Villum Foundation�s VKR Centre of Excellence �NAMEC� (Contract No. 65286) and Young Investigator Programme (Project No. VKR023125).

\section*{Author contributions}
G.A.B. and M.R.V. contributed equally to this work. This quadratic measurement research program was conceived by M.R.V. with refinements from G.A.B. and W.P.B.. The optomechanical evanescent coupling setup was designed by G.A.B. and W.P.B. with later input from M.R.V.. G.A.B. was the main driving force behind building the experiment and performing the data analysis with important input from M.R.V. and W.P.B.. Micro-fabrication of the SiN nanostring mechanical resonators was performed by P.E.L, S.S., and A.B.. This manuscript was written by M.R.V. and G.A.B. with important contributions from W.P.B.. Overall laboratory leadership was provided by W.P.B. and substantial supervision for this project was performed by M.R.V..

\section*{Additional information}
The authors declare no competing financial interests.

\end{document}